\documentstyle[psfig]{mn}
\newif\ifAMStwofonts

\newcommand{\simgt}{\lower.5ex\hbox{$\; \buildrel > \over \sim \;$}}
\newcommand{\simlt}{\lower.5ex\hbox{$\; \buildrel < \over \sim \;$}}

\newcommand{\sbkt}[1]{\left(#1\right)}

\newcommand{\msun}{\,{\rm M}_{\odot}}

\def\gsim{\mathrel{\mathpalette\Oversim>}}
\def\Oversim#1#2{\lower0.5ex\vbox{\baselineskip0pt\lineskip0pt%
            \lineskiplimit0pt\ialign{%
          $\mathsurround0pt #1\hfil##\hfil$\crcr#2\crcr\sim\crcr}}}

\ifoldfss
  \ifCUPmtlplainloaded \else
    \NewTextAlphabet{textbfit} {cmbxti10} {}
    \NewTextAlphabet{textbfss} {cmssbx10} {}
    \NewMathAlphabet{mathbfit} {cmbxti10} {} 
    \NewMathAlphabet{mathbfss} {cmssbx10} {} 
  \fi
  \ifAMStwofonts
    \ifCUPmtlplainloaded \else
      \NewSymbolFont{upmath} {eurm10}
      \NewSymbolFont{AMSa} {msam10}
      \NewMathSymbol{\upi}     {0}{upmath}{19}
      \NewMathSymbol{\umu}     {0}{upmath}{16}
      \NewMathSymbol{\upartial}{0}{upmath}{40}
      \NewMathSymbol{\leqslant}{3}{AMSa}{36}
      \NewMathSymbol{\geqslant}{3}{AMSa}{3E}

      \let\leq=\leqslant 
       
    \fi
  \fi
\fi 

\ifnfssone
  \newmathalphabet{\mathit}
  \addtoversion{normal}{\mathit}{cmr}{m}{it}
  \addtoversion{bold}{\mathit}{cmr}{bx}{it}
  \newmathalphabet{\mathbfit} 
  \addtoversion{normal}{\mathbfit}{cmr}{bx}{it}
  \addtoversion{bold}{\mathbfit}{cmr}{bx}{it}
  \newmathalphabet{\mathbfss} 
  \addtoversion{normal}{\mathbfss}{cmss}{bx}{n}
  \addtoversion{bold}{\mathbfss}{cmss}{bx}{n}
  \ifAMStwofonts
    \ifCUPmtlplainloaded \else
      \UseAMStwoboldmath
      \makeatletter
      \new@mathgroup\upmath@group
      \define@mathgroup\mv@normal\upmath@group{eur}{m}{n}
      \define@mathgroup\mv@bold\upmath@group{eur}{b}{n}
      \edef\UPM{\hexnumber\upmath@group}
      \new@mathgroup\amsa@group
      \define@mathgroup\mv@normal\amsa@group{msa}{m}{n}
      \define@mathgroup\mv@bold\amsa@group{msa}{m}{n}
      \edef\AMSa{\hexnumber\amsa@group}
      \makeatother
      \mathchardef\upi="0\UPM19
      \mathchardef\umu="0\UPM16
      \mathchardef\upartial="0\UPM40
      \mathchardef\leqslant="3\AMSa36
      \mathchardef\geqslant="3\AMSa3E

      \let\leq=\leqslant 

    \fi
  \fi
\fi 

\ifnfsstwo
  \DeclareMathAlphabet{\mathbfit}{OT1}{cmr}{bx}{it}
  \SetMathAlphabet\mathbfit{bold}{OT1}{cmr}{bx}{it}
  \DeclareMathAlphabet{\mathbfss}{OT1}{cmss}{bx}{n}
  \SetMathAlphabet\mathbfss{bold}{OT1}{cmss}{bx}{n}
  \ifAMStwofonts
    \ifCUPmtlplainloaded \else
      \DeclareSymbolFont{UPM}{U}{eur}{m}{n}
      \SetSymbolFont{UPM}{bold}{U}{eur}{b}{n}
      \DeclareSymbolFont{AMSa}{U}{msa}{m}{n}
      \DeclareMathSymbol{\upi}{0}{UPM}{"19}
      \DeclareMathSymbol{\umu}{0}{UPM}{"16}
      \DeclareMathSymbol{\upartial}{0}{UPM}{"40}
      \DeclareMathSymbol{\leqslant}{3}{AMSa}{"36}
      \DeclareMathSymbol{\geqslant}{3}{AMSa}{"3E}

      \let\leq=\leqslant 

    \fi
  \fi
\fi 

\ifCUPmtlplainloaded \else
  \ifAMStwofonts \else 
    \def\upi{\pi}
    \def\umu{\mu}
    \def\upartial{\partial}
  \fi
\fi

\title[Radiation-Hydrodynamical Collapse in the UV background]{Radiation-Hydrodynamical Collapse of Pregalactic Clouds in the
Ultraviolet Background}
\author[T. Kitayama et al.]
       {T. Kitayama$^{1}$, Y. Tajiri$^{2}$, M. Umemura$^{3}$,
H. Susa$^{3}$ and S. Ikeuchi$^{4}$ \\
$^{1}$ Department of Physics, Tokyo Metropolitan University, 
Hachioji, Tokyo 192-0397, Japan\\
$^{2}$ Institute of Physics,  University of Tsukuba, Tsukuba 305-8577, 
Japan \\
$^{3}$ Center for Computational Physics, University of Tsukuba, 
Tsukuba 305-8577, Japan\\
$^{4}$ Department of Physics, Nagoya University, Chikusa-ku, Nagoya
464-8602, Japan}

\pagerange{\pageref{firstpage}--\pageref{lastpage}}
\pubyear{2000}

\begin{document}

\maketitle

\label{firstpage}

\begin{abstract}
To elucidate the effects of the UV background radiation on the
collapse of pregalactic clouds, we implement a
radiation-hydrodynamical calculation, combining one-dimensional
spherical hydrodynamics with an accurate treatment of the radiative
transfer of ionizing photons. Both absorption and scattering of UV
photons are explicitly taken into account. It turns out that a gas
cloud contracting within the dark matter potential does not settle
into hydrostatic equilibrium, but undergoes run-away collapse even
under the presence of the external UV field. The cloud center is shown
to become self-shielded against ionizing photons by radiative transfer
effects before shrinking to the rotation barrier. Based on our
simulation results, we further discuss the possibility of H$_2$
cooling and subsequent star formation in a run-away collapsing core.
The present results are closely relevant to the survival of
subgalactic Population III objects as well as to metal injection into
intergalactic space.
\end{abstract}


\section{Introduction}

It is widely recognized that the Ultraviolet (UV) background
radiation, inferred from the proximity effect of Ly$\alpha$ absorption
lines in QSO spectra (e.g., Bajtlik, Duncan \& Ostriker 1988; Bechtold
1994; Giallongo et al. 1996), is likely to exert a significant
influence upon the collapse of pregalactic clouds and consequently the
formation of galaxies. Several authors have argued that the formation
of subgalactic objects is suppressed via photoionization and
photoheating caused by the UV background (Umemura \& Ikeuchi 1984;
Ikeuchi 1986; Rees 1986; Bond, Szalay \& Silk 1988; Efstathiou 1992;
Babul \& Rees 1992; Zhang, Anninos \& Norman 1995; Thoul \& Weinberg
1996). Final states of the photoionized clouds, however, are still
unclear, because most of these studies assume optically-thin media,
which precludes us from the correct assessment of self-shielding
against the external UV fields.  Self-shielding not only plays an
important role in the thermal and dynamical evolution, but is also
essential for the formation of hydrogen molecules (H$_2$) which
control the star formation in metal poor environments such as
primordial galaxies.
Several attempts have recently been made to take account of the
radiative transfer of ionizing photons, adopting for instance a pure
absorption approximation (Kepner, Babul \& Spergel 1997; Kitayama \&
Ikeuchi 2000, hereafter KI), the photon conservation method (Abel,
Norman \& Madau 1999), and the full radiative transfer treatment
(Tajiri \& Umemura 1998, hereafter TU; Barkana \& Loeb 1999; Susa \&
Umemura 2000).  As for the dynamical states, Kepner, Babul \& Spergel
(1997) and Barkana \& Loeb (1999) have considered {\it hydrostatic}
equilibria of spherical clouds within virialized dark halos. KI have
explored the {\it hydrodynamical} evolution of a spherical system
composed of dark matter and baryons.  The full radiative transfer
treatment, however, has not hitherto been incorporated with the
hydrodynamics of spherical pregalactic clouds.

In this paper, we attempt to implement an accurate
radiation-hydrodynamical (RHD) calculation on the evolution of
spherical clouds exposed to the UV background, solving simultaneously
the radiative transfer of photons and the gas hydrodynamics.  Our
goals are to predict at a physically reliable level final states of
photoionized clouds and also to assess accurately self-shielding
against the UV background that is essential for H$_2$ cooling and the
subsequent star formation.  Throughout the present paper, we assume
the density parameter $\Omega_0=1$, the Hubble constant $h=H_0$/(100
km s$^{-1}$ Mpc$^{-1})=0.5$, and the baryon density parameter
$\Omega_{\rm b} = 0.1$.

\section{Model}
\label{sec:model}

A pregalactic cloud is supposed to be a mixture of baryonic gas and
dark matter with the mass ratio of $\Omega_{\rm b} : \Omega_0 -
\Omega_{\rm b} = 1 : 9$. The numerical scheme for the spherical
Lagrangian dynamics of two-component matter follows the method
described in Thoul \& Weinberg (1995) and KI.  At each time-step, the
radiative transfer is solved with the method devised by TU, in which
both absorption and emission (scattering) of ionizing photons are
explicitly taken into account.  We assume for simplicity that the
baryonic gas is composed of pure hydrogen.  Neglecting helium causes
only a minor effect in the ionization degree less than about the order
of 10\% (Osterbrock 1989; Nakamoto, Susa \& Umemura 1998). We further
assume ionization equilibrium among photoionization, collisional
ionization, and recombination.  This assumption is well justified in
the present analysis (see \S 2.2 of KI for discussion).  The number of
mass shells is $N_{\rm b}=200$ for baryonic gas and $N_{\rm d}= 2000$
for dark matter. At each radial point, angular integration of the
radiative transfer equation is done over at least 20 bins in $\theta =
0 -\pi$, where $\theta$ is the angle between the light ray and the
radial direction. This is achieved by handling $300 - 700$ impact
parameters for light rays. The radiation field and the ionization
states in the cloud interior are solved iteratively until the HI
fraction, $X_{\rm HI}$, in each mesh converges within an accuracy of
1\%.

The external UV field is presumed to be isotropic and to have a
power-law spectrum:
\begin{equation}
  J(\nu) = J_{21} \sbkt{\frac{\nu}{\nu_{\rm HI}}}^{-\alpha}
        \times 10^{-21} \mbox{erg s$^{-1}$ cm$^{-2}$ str$^{-1}$
        Hz$^{-1}$},
\label{eq:uvb}
\end{equation}
where $J_{21}$ is the intensity at the Lyman edge of hydrogen
($h\nu_{\rm HI} = 13.6$eV) and $\alpha$ is the spectral index. We
consider two typical cases for $\alpha$, i.e., $\alpha=1$
representative for black hole accretion and $\alpha=5$ for stellar UV
sources. Observations of the proximity effect in the Ly$\alpha$ forest
suggest $J_{21} = 10^{\pm 0.5}$ at $z=1.7-4.1$ (e.g., Bajtlik, Duncan
\& Ostriker 1988; Bechtold 1994; Giallongo et al. 1996), but its value
is still uncertain at other redshifts. In what follows we give the
onset of the UV background to be at $z_{\rm UV} =20$ and study the
following two cases for $J_{21}$:
\begin{enumerate}
\item  Constant UV 
\begin{equation}
  \hspace*{1cm}  J_{21} = 1  \hspace*{2.5cm} z \leq z_{\rm UV}, 
\end{equation}
\item  Evolving UV 
\begin{equation}
 \hspace*{1cm} J_{21} = \left\{\begin{array}{ll}
       \exp[-(z-5)] &  5 \leq z \leq z_{\rm UV} \\
      1 &  3 \leq z \leq 5 \\ 
      \sbkt{\frac{1+z}{4}}^4 &  0 \leq z \leq 3. \\ 
     \end{array} \right.
\label{eq:uvevl}
\end{equation}
\end{enumerate}
The form of the UV evolution at $z>5$ in (ii) is roughly consistent
with the results of recent models for the reionization of the universe
(e.g., Ciardi et al 2000; Umemura, Nakamoto \& Susa 2000).


We start the simulations when the overdensity of a cloud is still in
the linear regime, adopting the initial and boundary conditions
described in KI.  The initial overdensity profile is $\delta_{\rm
i}(r) = \delta_{\rm i}(0)\sin(kr)/kr$, where $k$ is the comoving wave
number, and the central overdensity $\delta_{\rm i}(0)$ is fixed at
0.2.  The outer boundary is taken at the first minimum of $\delta_{\rm
i}(r)$, i.e., $kr = 4.4934$, within which the volume averaged
overdensity $\bar{\delta}(<r)$ vanishes. Following Haiman, Thoul \&
Loeb (1996), the characteristic mass of a cloud $M_{\rm cloud}$ is
defined as the baryon mass enclosed within the first zero of
$\delta_{i}(r)$, i.e., $kr = \pi$. Collapse redshift $z_{\rm c}$ is
defined as the epoch at which $M_{\rm cloud}$ would collapse to the
center in the absence of thermal pressure.  Circular velocity $V_{\rm
c}$ and virial temperature $T_{\rm vir}$ are related to $z_{\rm c}$
and $M_{\rm cloud}$ via usual definitions:
\begin{equation}
 V_c = 15.9 \sbkt{\frac{M_{\rm cloud} \Omega_0/\Omega_b}
{10^9 h^{-1} \msun}}^{1/3} (1+z_c)^{1/2} ~\mbox{  km s$^{-1}$}, 
\end{equation}
\begin{equation}
T_{\rm vir} = 9.09 \times 10^3  \sbkt{\frac{\mu}{0.59}}
\sbkt{\frac{M_{\rm cloud} \Omega_0/\Omega_b}
{10^9 h^{-1} \msun}}^{2/3} (1+z_c) ~\mbox{  K}, 
\label{eq:vc}
\end{equation}
where $\mu$ is the mean molecular weight in units of the proton mass
$m_{\rm p}$.

The collapse of a gas shell is traced until it reaches the rotation
radius specified by the dimensionless spin parameter;
\begin{equation}
r_{\rm rot}= 0.05 \sbkt{\frac{\Omega_{\rm b}/\Omega_0}{0.1}}^{-1}
\left( \lambda_{\rm ta} \over 0.05 \right)^2 r_{\rm ta},
\label{eq:rrot}
\end{equation}
where $r_{\rm ta}$ is the turnaround radius of the gas shell, and we
adopt a median for the spin parameter, $\lambda_{\rm ta}=0.05$
(Efstathiou \& Jones 1979; Barns \& Efstathiou 1987; Warren et
al. 1992). Below the size given by (\ref{eq:rrot}), the system 
would attain rotational balance and forms a disk eventually.

\section{Results}

\subsection{Significance of radiative transfer}

To demonstrate the significance of coupling the radiative transfer of
ionizing photons with hydrodynamics, we first present in Fig.~1
estimations for relevant timescales at the center of a uniform static
cloud with circular velocity $V_{\rm c}$ and collapse epoch $z_{\rm
c}$, where the external UV is specified by $J_{21}$ and a spectral
index $\alpha$.  The photoionization timescale $t_{\rm ion}$ and the
photoheating timescale $t_{\rm heat}$ are compared to the dynamical
timescale $t_{\rm dyn}$, either by solving the radiative transfer of
UV photons through the cloud or by just assuming the optically thin
medium.  Wherever necessary, we have adopted the temperature
$T=10^4$K, the proton number density $n_{\rm H}=n_{\rm H}^{\rm vir} =
5.0 \times 10^{-5} (\Omega_{\rm b} h^2/0.025) (1+z_{\rm c})^3$
cm$^{-3}$, the mass density $\rho= m_p n_{\rm H} \Omega_0/\Omega_b$,
and the radius $R=R_{\rm vir}=64 (V_{\rm c}/30 \mbox{km s$^{-1}$})
(\Omega_0 h^2/0.25)^{-1/2} (1+z_{\rm c})^{-3/2}$ kpc. In the
low-density limit, the contours of $t_{\rm ion}=t_{\rm dyn}$ and
$t_{\rm heat}=t_{\rm dyn}$ both approach asymptotically the optically
thin case $J_{21} \propto \sqrt{n_{\rm H}^{\rm vir}} \propto
(1+z_c)^{3/2}$. The differences originated from the cloud sizes are
rather small compared to those arisen when one incorporates the
radiative transfer or not.


Fig.~1 predicts that the significance of radiative transfer effects
increases with increasing redshift (i.e., increasing cloud density)
and decreasing $J_{21}$ for a given $\alpha$. Under the UV evolution
given in equation (\ref{eq:uvevl}), for example, $t_{\rm dyn}< t_{\rm
ion}$ and $t_{\rm dyn} < t_{\rm heat}$ are satisfied at the center of
a cloud with $V_c=30$km s$^{-1}$ at $z_{\rm c}\simgt 10$, and the
cloud is shielded against the external radiation in terms of
photoionization as well as photoheating. At $6 \simlt z_{\rm c}\simlt
10$, $t_{\rm heat} < t_{\rm dyn} < t_{\rm ion}$ is achieved,
indicating that the cloud center is heated but not ionized by the UV
radiation (Gnedin \& Ostriker 1997; KI). Contrastively, both $t_{\rm
ion}$ and $t_{\rm heat}$ become shorter than $t_{\rm dyn}$ at $z_{\rm
c} < 6$ and the cloud is likely to evolve in a similar fashion to the
optically-thin case.  We can see qualitatively the same relations for
$\alpha = 5$, except that shielding of ionization and heating occurs
almost simultaneously at lower redshift.  It should be noted that the
above estimations are made for a virialized cloud using the UV
intensity only at $z_c$. As shown in forthcoming sections, the central
density of a collapsing cloud actually continues to ascend to above
$n_{\rm H}^{\rm vir}$ and the radiative transfer effects can become
important even at low redshifts.  In addition, possible changes of the
UV intensity during the dynamical growth of a cloud can also affect
its ionization structure.

\subsection{Dynamical evolution}

Fig.~\ref{fig:trajc} shows the dynamical evolution of a cloud with
$V_{\rm c}=29$km s$^{-1}$ under the constant UV background at $z < z_{\rm
UV}=20$. This cloud would collapse at $z_{\rm c}=3$ if there were no
UV background. In practice, the cloud turns around and contracts in
the inner parts, while it continues to expand in the outer envelope.
For a hard UV spectrum with $(J_{21},\alpha) = (1,1)$, the gas is
ionized and heated up to $T \sim 10^4$K promptly at the onset of the
UV background. For a soft UV spectrum with $(J_{21},\alpha) = (1,5)$,
the cloud center is kept self-shielded against the external field and
the temperature ascends more gradually.

For comparison, results of the optically-thin and pure absorption
calculations are also presented in Fig.~\ref{fig:trajc}. The pure
absorption case is based on the analytical formalism described in
KI. The cloud evolution is altered in no small way by different
treatments of the UV radiation. Under the optically-thin assumption,
the cloud is completely prohibited from collapsing for
$(J_{21},\alpha) = (1,1)$, and from being self-shielded for
$(J_{21},\alpha) = (1,5)$.  The pure absorption approximation leads to
accelerating the collapse of the central core and to underestimating
photoionization and photoheating.

Fig.~\ref{fig:prof} exhibits the radial profiles of the same clouds at
$z=3$ for $(J_{21},\alpha) = (1,5)$. The inner part of a cloud does
not settle into hydrostatic equilibrium, but rather undergoes
isothermal run-away collapse and the density profile follows the
self-similar solution of Bertschinger (1985). The trend is regardless
of the treatment of the UV radiation transfer, because the temperature
of thermal equilibrium in optically-thin media comes close to $10^4$K
in the high density limit. The present results demonstrate that the
approximation of hydrostatic equilibrium employed in previous analyses
(Kepner, Babul \& Spergel 1997; Barkana \& Loeb 1999) breaks down for
the final states.

The ionization structure is quite different depending on whether one
incorporates the radiative transfer or not. In an optically-thin cloud,
the ionization degree changes merely according to the ionization
parameter, $n_{\rm H}/J_{21}$, whereas the radiative transfer effects
produce a self-shielded neutral core (Fig.~\ref{fig:prof}c).
Fig.~\ref{fig:prof}(d) further indicates that the UV heating rate is
reduced significantly in the self-shielded region, with an increasing
contribution of scattered photons to the total heating
rate. Implications of the present results on H$_2$ cooling [thin lines
in panel (d)] will be discussed in detail in Sec \ref{sec:sf}.

Effects of the evolution of the UV background are illustrated in
Fig.~\ref{fig:traje}. Due to very low UV intensity at
high redshift, the whole cloud is kept self-shielded in terms of both
photoionization and photoheating at $z \simgt 12$. For a cloud with
relatively high $z_c$ ($z_c=4.8$, Fig.~\ref{fig:traje}a), the cloud
center begins to contract before the penetration of the external
UV. Hence, the dynamical evolution closely coincides with that without
the UV background.  As the virial temperature of the cloud is $T_{\rm
vir} \sim 3\times 10^4$K, the could center is shock-heated to above
$10^4$K and cools via atomic cooling. On the other hand, a cloud with
low $z_c$ ($z_c=0.5$, Fig.~\ref{fig:traje}b) is once photoionized to
the similar level to the constant UV case at $3 < z < 5$, but is able
to collapse as the UV intensity drops at lower redshifts.

\subsection{Criteria for collapse} 

Figs \ref{fig:vzc} and \ref{fig:vze} summarize the results of present
calculations for a variety of initial conditions on a $V_{\rm c} -
z_{\rm c}$ plane. As our simulations assume ionization equilibrium and
only incorporate atomic cooling, they can be most securely applied to
a cloud once heated to above $10^4$K either by shock or by UV photons
during the course of its evolution. In contrast, evolution of lower
temperature systems is still tentative and may be altered once cooling
by molecular hydrogen is explicitly taken into account.  These figures
thus distinguish ``high temperature clouds'' (circles) defined as
those photoheated to $ > 10^4$K or those with $T_{\rm vir} > 10^4$K,
and the other ``low temperature clouds'' (triangles). Each of these
populations are further classified by open and filled symbols
depending on whether or not they collapse to the rotation barrier
within the present age of the universe.

It is obvious that the UV background prohibits small clouds from
collapsing even if the transfer of ionizing photons is considered.
Under a constant UV flux (Fig.~\ref{fig:vzc}), threshold circular
velocity for the collapse gradually increases with decreasing
redshift, i.e., decreasing cloud density. The threshold velocity also
decreases with increasing photon spectral index $\alpha$ because of
the smaller number of high energy photons. At $z_c \simgt 5$, the
threshold falls below $10^4$K, because the cloud center starts to
contract before the onset of the UV background and remains impervious
to the external photons. In the presence of the UV evolution
(Fig.~\ref{fig:vze}), the threshold velocity increases sharply at $z_c
\simlt 3$ and drops slightly at $z_c \simlt 1$.  The central
temperature of a cloud denoted by an open triangle can reach $\sim
10^4$K by adiabatic compression and the collapse is promoted by atomic
cooling. A cloud denoted by a filled triangle fails to collapse
because of the lack of coolant at $T<10^4$K in our simulations. As
will be discussed in Sec \ref{sec:sf}, however, evolution of these
``low temperature clouds'' may be modified by H$_2$ cooling.

KI have suggested that the threshold for collapse is roughly
determined by the balance between the gravitational force and the
thermal pressure gradient when the gas is maximally exposed to the
external UV flux. To confirm this, we plot in the same figures a
relation $T_{\rm vir} = T_{\rm eq}^{\rm max}$, where $T_{\rm eq}^{\rm
max}$ is the maximum equality temperature defined semi-analytically as
follows. Firstly, given the initial overdensity profile, the collapse
of a spherical perturbation is approximated by the self-similar
solution of Bertschinger (1985). Secondly, at an arbitrary stage of
the collapse, one can compute the equality temperature at which
radiative cooling balances photoheating in the optically thin limit,
using the central density deduced from the self-similar solution and
the UV intensity assumed in the simulation.  Finally, $T_{\rm eq}^{\rm
max}$ is set equal to the maximum of such temperature.  For the gas
exposed to a constant UV flux from the linear regime, $T_{\rm eq}^{\rm
max}$ is attained essentially at turn-around. If a cloud has $T_{\rm
vir}$ above $T_{\rm eq}^{\rm max}$, it is likely to be gravitationally
unstable against gas pressure.

Figs \ref{fig:vzc} and \ref{fig:vze} show that the above criteria
agree reasonably well with our simulation results based on the
accurate treatment of the radiative transfer. This is because the
dynamical evolution is basically regulated by the Jeans criterion when
a cloud is heated up to $\sim 10^4$K by the UV background before
contraction. At high redshifts ($z_c \simgt 6$ in Fig.~\ref{fig:vzc}
and $z_c \simgt 3$ in Fig.~\ref{fig:vze}), however, the cloud center
is strongly self-shielded from early stages and the approximation of
the optically thin limit breaks down. The Jeans scale at these epochs
is reduced below the relation $T_{\rm vir} = T_{\rm eq}^{\rm max}$.

\section{Implications for Galaxy Formation in the UV Background}
\label{sec:sf}

The present RHD calculations give an accurate prediction for the
suppression of pregalactic collapse due to the UV background, which
has been one of the primary concerns from the viewpoint of galaxy
formation (Umemura \& Ikeuchi 1984; Ikeuchi 1986; Rees 1986; Bond,
Szalay \& Silk 1988; Efstathiou 1992; Babul \& Rees 1992; Zhang,
Anninos \& Norman 1995; Thoul \& Weinberg 1996; KI). What is of
additional significance in this context is the subsequent formation of
stars in collapsing clouds. In order for stars to form, a cloud needs
to be cooled down to below $10^4$K by hydrogen molecules, because they
are the only coolant in metal-deficient gas (e.g., Peebles \& Dicke
1968; Matsuda, Sato \& Takeda 1969; Tegmark et al. 1997). H$_2$
cooling is a two-body collision process (e.g., Hollenbach \& Mckee
1989; Galli \& Palla 1998), while the photoheating rate is in
proportion to the density. The potentiality of H$_2$ cooling is
therefore enhanced with increasing density. In this respect, run-away
collapse should provide favorable situations for H$_2$ cooling.

Based on our simulation results presented in the previous section, we
further investigate the possibility of star formation in a collapsing
core. We suppose that cooling below $10^4$K becomes efficient if the
following two conditions are both satisfied; 1) photodissociation of
molecular hydrogen by the UV photons in the Lyman-Werner bands at
$11.26 - 13.6$keV (e.g., Stecher \& Williams 1967) is less important
than other H$_2$ destruction processes, and 2) H$_2$ cooling overtakes
UV photoheating. The first condition is fulfilled for clouds heated up
to $>10^4$K, because destruction of H$_2$ is dominated by collisional
dissociation insofar as $T\gsim 2000$K (Corbelli, Galli \& Palla
1997). More specifically, requiring that the timescale of H$_2$
dissociation via collisions with H$^+$ (reaction 12 of Shapiro \& Kang
1987) is shorter than that of photodissociation (Draine \& Bertoldi
1996; Omukai \& Nishi 1999) in the conservative optically thin limit,
we have for the electron density
\begin{equation}
n_{\rm e} > 4.7 \times 10^{-5} \sbkt{\frac{13.6}{12.4}}^{\alpha}
 J_{21} \exp(21200\mbox{ K}/T) \mbox{~~cm$^{-3}$},
\label{eq:sol}
\end{equation}
which is amply satisfied at the collapsing core of our simulated
clouds. The second condition depends on the amount of H$_2$ formed. In
the absence of the external UV field, the H$_2$ abundance in the
metal-deficient postshock layer
converges roughly to $X_{\rm H2}\approx 10^{-3}$ (e.g., Shapiro \&
Kang 1987; Ferrara 1998; Susa et al. 1998). Under the UV field,
$X_{\rm H2}\approx 10^{-3}$ is also achieved if photoheating is
strongly attenuated by self-shielding, while the abundance is reduced
down to $X_{\rm H2}\approx 10^{-6}$ in the case of weak attenuation of
photoheating (Kang \& Shapiro 1992; Corbelli et al. 1997; Susa \&
Umemura 2000; Susa \& Kitayama 2000). It is also likely that the H$_2$
abundance depends on the spectrum of the impinging radiation and the
details of the ionizing structure of the medium (e.g., Kang \& Shapiro
1992; Ciardi et al. 2000).  Leaving such complexities elsewhere
(Kitayama et al. in preparation), we make crude estimations for the
cooling rate using the H$_2$ cooling function of Hollenbach \& Mckee
(1989) and assuming $X_{\rm H2}=10^{-3}$ in the full transfer and pure
absorption cases, and $X_{\rm H2}=10^{-6}$ in the optically thin case.

Fig.~\ref{fig:prof}(d) shows that H$_2$ cooling greatly overwhelms
photoheating in the self-shielded neutral core for $X_{\rm
H2}=10^{-3}$. It turns out that what enables H$_2$ cooling to be
effective is essentially the attenuation of photoheating rather than
the final abundance of H$_2$ molecules. In fact, in the strongly
self-shielded core, H$_2$ cooling remains to be dominant even if the
abundance is reduced to a level similar to the optically-thin case,
i.e., $X_{\rm H2}=10^{-6}$ (the H$_2$ cooling rate is roughly
proportional to $X_{\rm H2}$ at $T \simlt 10^4$ K). We have further
checked that this is the case with every collapsed cloud plotted as an
open circle in Figs \ref{fig:vzc} and \ref{fig:vze}.  Under the
optically-thin approximation, on the other hand, UV heating is much
stronger than H$_2$ cooling, prohibiting the cooling below $10^4$K.

While the above arguments support the possibility of H$_2$ cooling
during the collapse of ``high temperature clouds'' (open circles in
Figs \ref{fig:vzc} and \ref{fig:vze}), the situation is rather
intricate for ``low temperature clouds'' (triangles in the same
figures).  Once incorporated formation and destruction of molecular
hydrogen explicitly in our calculations, these clouds would be able to
collapse as long as the external UV flux is negligible or very weak.
For somewhat stronger UV flux, photodissociation may operate
efficiently to suppress H$_2$ formation and to prohibit the collapse
(Haiman, Rees \& Loeb 1996, 1997; Haiman, Abel \& Rees 2000; Ciardi et
al. 2000). Alternatively, self-shielding induced by run-away collapse
may still enable a sufficient amount of H$_2$ to be formed for the
system to cool down to $\sim 100$K. These points will be investigated
thoroughly in our future work (Kitayama et al. in preparation).

In summary, self-shielding against the UV background is indispensable
for H$_2$ cooling and subsequent star formation to proceed. Run-away
collapse is likely to promote clouds with $T_{\rm vir} > 10^4$K to
cool down before collapsing to the rotation barrier given by equation
(\ref{eq:rrot}). These results are closely relevant to the formation
of dwarf galaxies at high redshifts as well as to metal injection into
intergalactic space (e.g., Nishi \& Susa 1999). Note that the above
conclusions do not conflict with those of several recent works (Haiman
Rees \& Loeb 1996, 1997; Ciardi et al. 2000; Haiman et al. 2000)
that the so called "radiative feedback" is operating in the early
universe and suppresses the collapse of some objects via
photodissociation of H$_2$ molecules. Current simulations are mainly
aimed to study the clouds once heated to above $10^4$K either by shock
or by UV photons, for which photodissociation has minor impacts on
H$_2$ formation compared to collisional dissociation
(eq.~[\ref{eq:sol}]). In addition, run-away collapse yields a highly
self-shielded core which is particularly favorable for H$_2$
formation. Photodissociation may be of greater significance in smaller
clouds and lower density regions.

Finally, we discuss the effects of the geometry of cloud evolution
upon H$_2$ cooling. Near and above the Jeans scale, clouds are
expected to contract spherically to a first order approximation and
result in run-away collapse as shown in the present paper.  Clouds far
above the Jeans scale, on the other hand, are likely to undergo
pancake collapse.  The pancakes would not end up with run-away
collapse, because the gravity of the sheet is readily overwhelmed by
the pressure force.  Susa \& Umemura (2000) have explored pregalactic
pancake collapse including the UV transfer and H$_2$ formation, and
shown that the collapsing pancakes bifurcate, depending upon the
initial masses, into less massive UV-heated ones and more massive
cooled ones. This is because the self-shielding in sheet collapse is
governed by the column density of a pregalactic cloud. To conclude,
star formation in pregalactic clouds in the UV background is strongly
regulated by the behaviour of collapse and the manner of the radiation
transfer.

\section*{Acknowledgments}
We thank Andrea Ferrara for helpful comments, Taishi Nakamoto for
discussions, and Susumu Inoue for careful reading of the
manuscript. This work is supported in part by Research Fellowships of
the Japan Society for the Promotion of Science for Young Scientists,
No. 7202 (TK) and 2370 (HS).


\label{lastpage}
\onecolumn
\begin{figure}
\begin{center}
   \leavevmode\psfig{figure=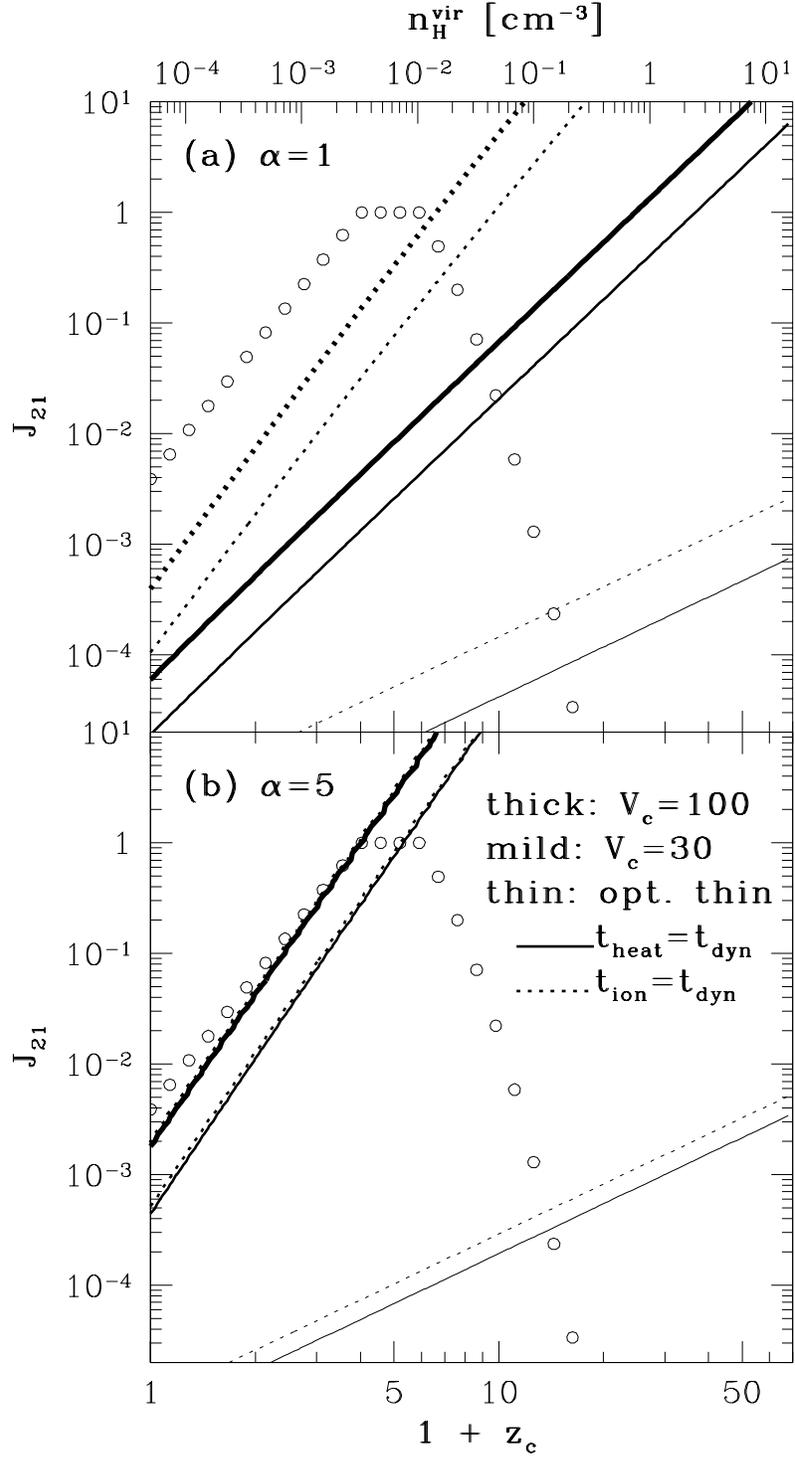,width=20cm}
\end{center}
\caption{$J_{21} - z_{\rm c}$ diagram on timescales relevant to the
evolution of a pregalactic cloud in the UV background with (a)
$\alpha=1$ and (b) $\alpha=5$. The photoionization time $t_{\rm ion}$,
the photoheating time $t_{\rm heat}$, and the dynamical time $t_{\rm
dyn}$ are evaluated at the center of a stationary uniform cloud with
$V_{\rm c}$ and $z_{\rm c}$ as described in the text. Lines indicate
the contours $t_{\rm ion}=t_{\rm dyn}$ (dotted) and $t_{\rm heat} =
t_{\rm dyn}$ (solid) for $V_{\rm c}=100$ km s$^{-1}$ (thick), 30 km
s$^{-1}$ (mildly thick), and in the optically thin limit (thin).  In
the region under these contours, $t_{\rm dyn}$ becomes shorter than
$t_{\rm ion}$ and $t_{\rm heat}$, respectively. For reference, the
track of the UV evolution given in equation
(\protect\ref{eq:uvevl}\protect) is marked by open circles.}
\label{fig:jz}
\end{figure}
\begin{figure}
\begin{center}
   \leavevmode\psfig{figure=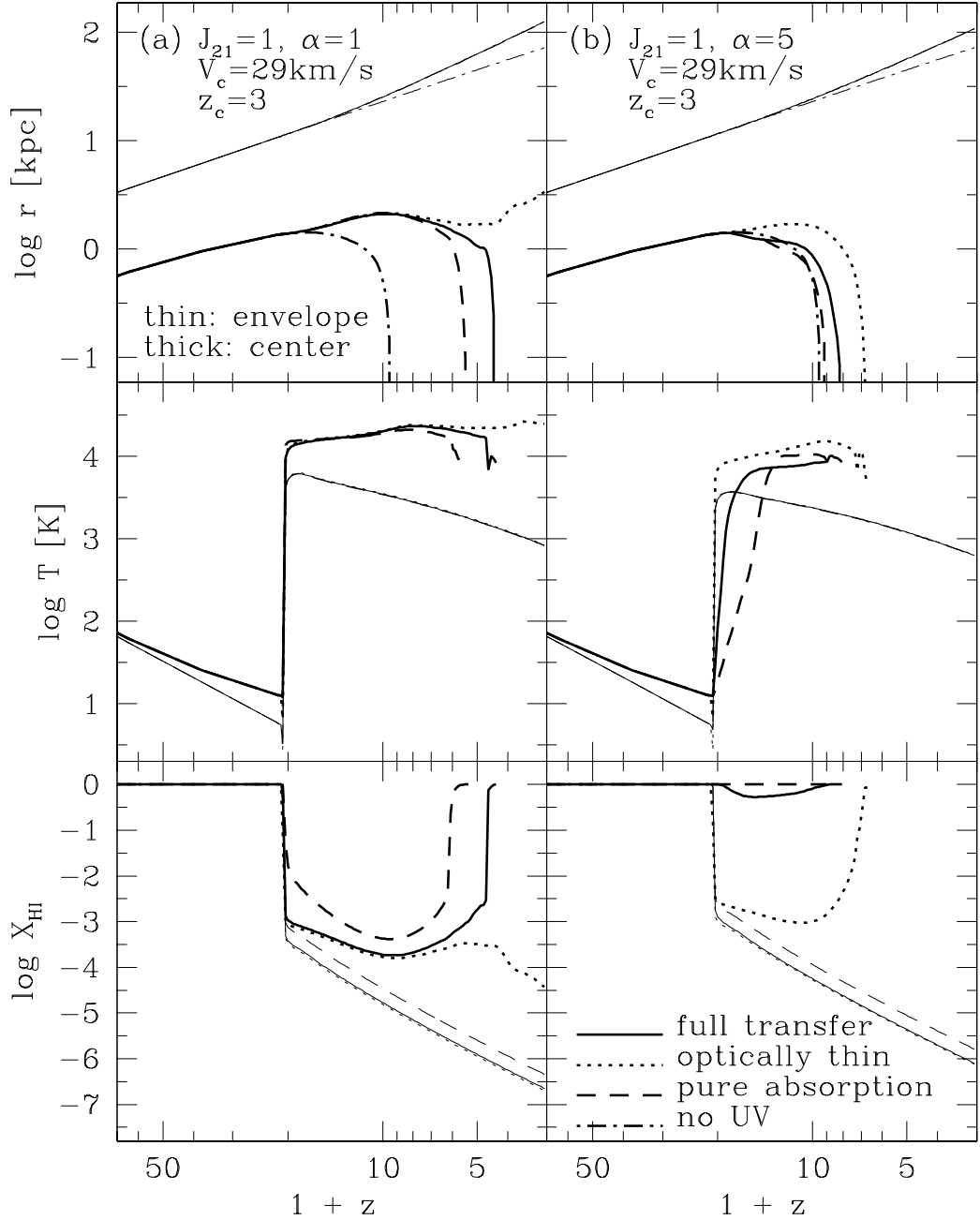,width=20cm}
\end{center}
\caption{Evolution of radius (top panels), temperature (middle
panels), and HI fraction (bottom panels) of gas shells enclosing 0.5\%
(thick lines) and 90\% (thin lines) of total gas mass in a cloud with
$V_{\rm c}=29$ km/s and $z_{\rm c}=3$ under the constant UV
background; (a) $(J_{21},\alpha)=(1,1)$ and (b) $(1,5)$. Different
line types indicate the full transfer (solid), optically-thin
(dotted), and pure absorption (dashed) cases, respectively. Also
plotted in the top panels are the no UV (dot-dashed) results.  }
\label{fig:trajc}
\end{figure}
\begin{figure}
\begin{center}
   \leavevmode\psfig{figure=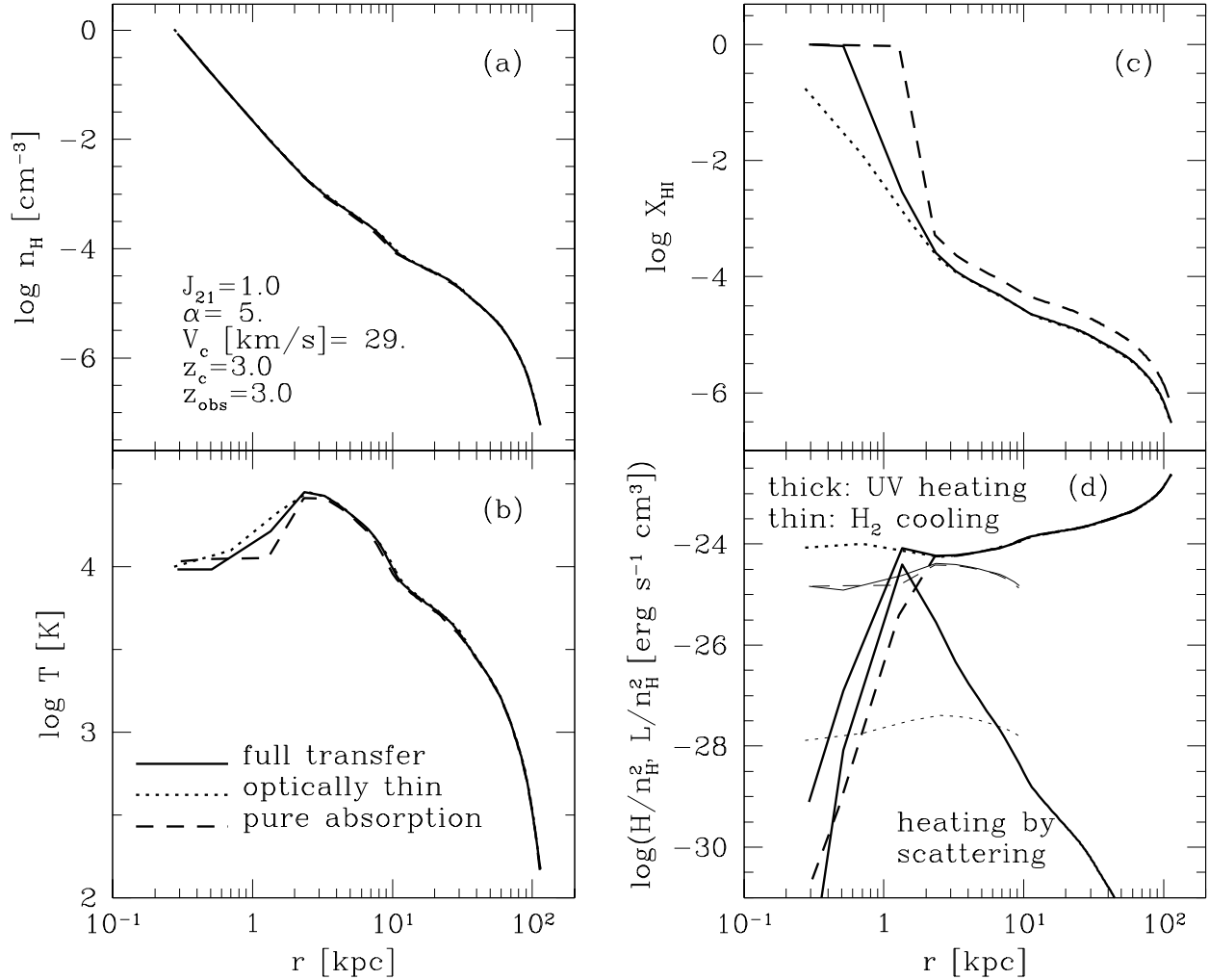,width=18cm}
\end{center}
\caption{Radial profiles at $z=3$ of (a) hydrogen density, (b)
temperature, (c) HI fraction, and (d) UV heating rate, for
$(J_{21},\alpha)=(1,5)$, $V_{\rm c}=29$ km/s, and $z_{\rm c}=3$.
Different line types indicate the full transfer (solid),
optically-thin (dotted), and pure absorption (dashed) cases,
respectively.  The heating rate caused by scattering in the full
transfer case is added to panel (d). Also plotted in (d) are
estimations for the H$_2$ cooling rate inside the shock front at $r
\sim 10$kpc, assuming $X_{\rm H2}=10^{-3}$ in the full transfer or
pure absorption cases, and $X_{\rm H2}=10^{-6}$ in the optically-thin
case.}
\label{fig:prof}
\end{figure}
\begin{figure}
\begin{center}
   \leavevmode\psfig{figure=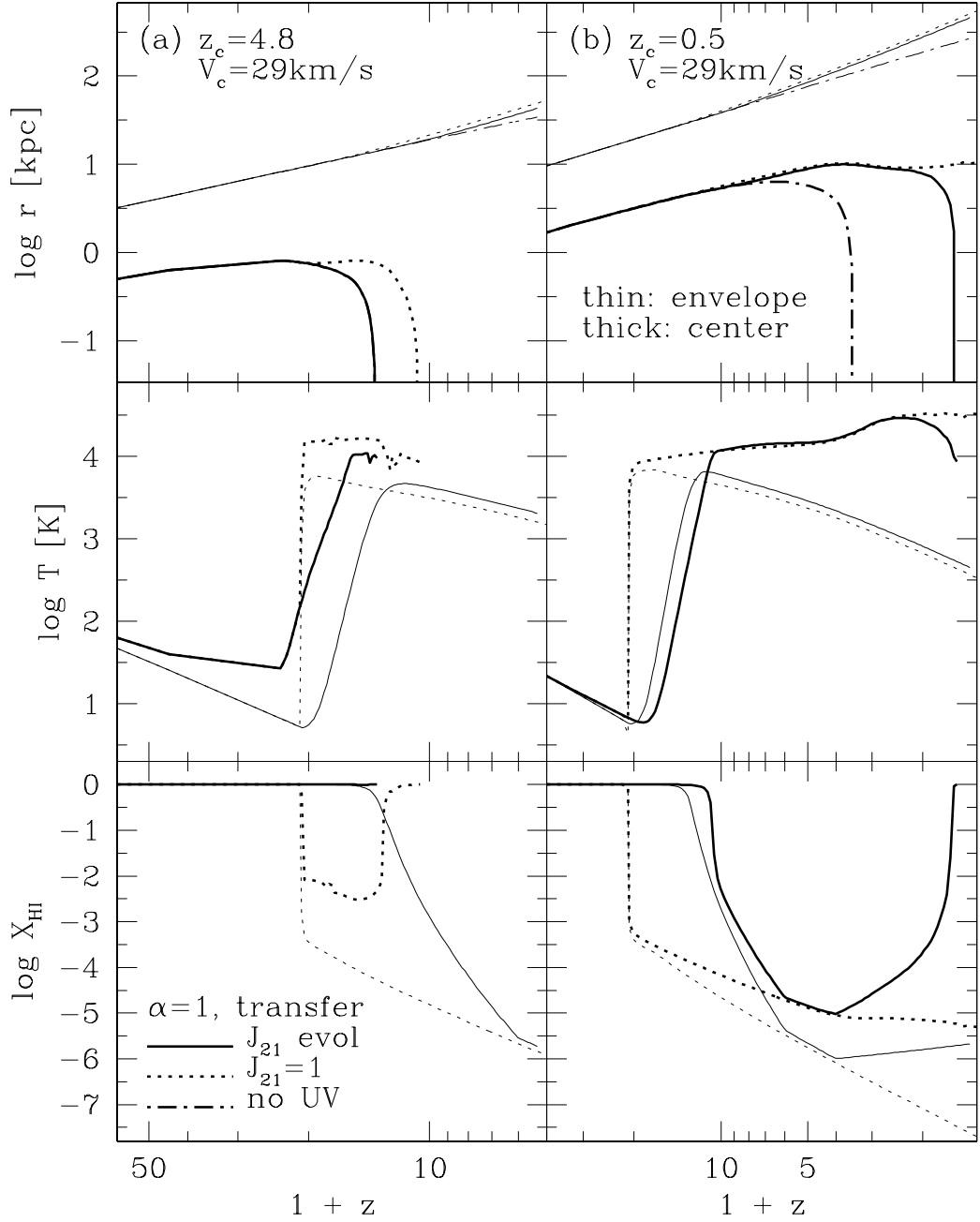,width=20cm}
\end{center}
\caption{Similar to Fig.~\protect\ref{fig:trajc}\protect~ except that
the effects of the UV evolution (solid lines) are illustrated in
comparison to the constant UV (dotted) and no UV (dot-dashed) cases;
(a) $z_{\rm c}=4.8$, $V_{\rm c}=29$ km/s, and (b) $z_{\rm c}=0.5$,
$V_{\rm c}=29$ km/s. Full transfer results are plotted assuming
$\alpha=1$.}
\label{fig:traje}
\end{figure}
\begin{figure}
\begin{center}
   \leavevmode\psfig{figure=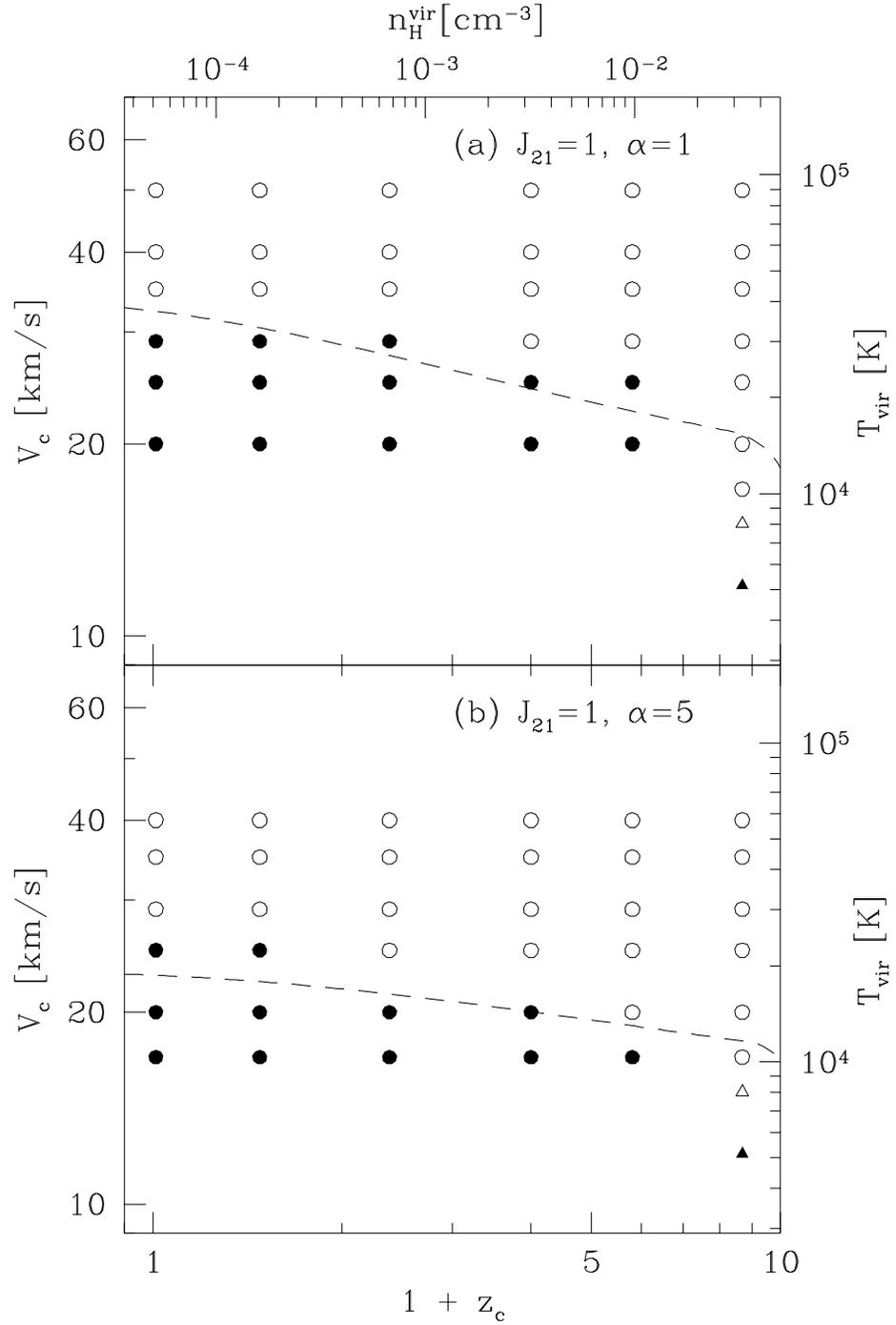,width=20cm}
\end{center}
\caption{$V_{\rm c} - z_{\rm c}$ diagram describing collapse and
self-shielding from our full-transfer RHD calculations; (a)
$(J_{21},\alpha)=(1,1)$, and (b) (1,5). Circles are the present secure
results for ``high temperature clouds'' and triangles are the
tentative results for ``low temperature clouds'' (see text for
definitions). Filled circles are clouds which are prohibited to
collapse due to the UV heating, and open circles are those which
undergo run-away collapse with $T_{\rm vir} > 10^4$K.  Filled
triangles are clouds which cannot collapse to the rotation barrier,
and open triangles are those which collapse due to atomic cooling.
Also plotted by dashed lines are the relation $T_{\rm vir} = T_{\rm
eq}^{\rm max}$ defined in the text.}
\label{fig:vzc}
\end{figure}
\begin{figure}
\begin{center}
   \leavevmode\psfig{figure=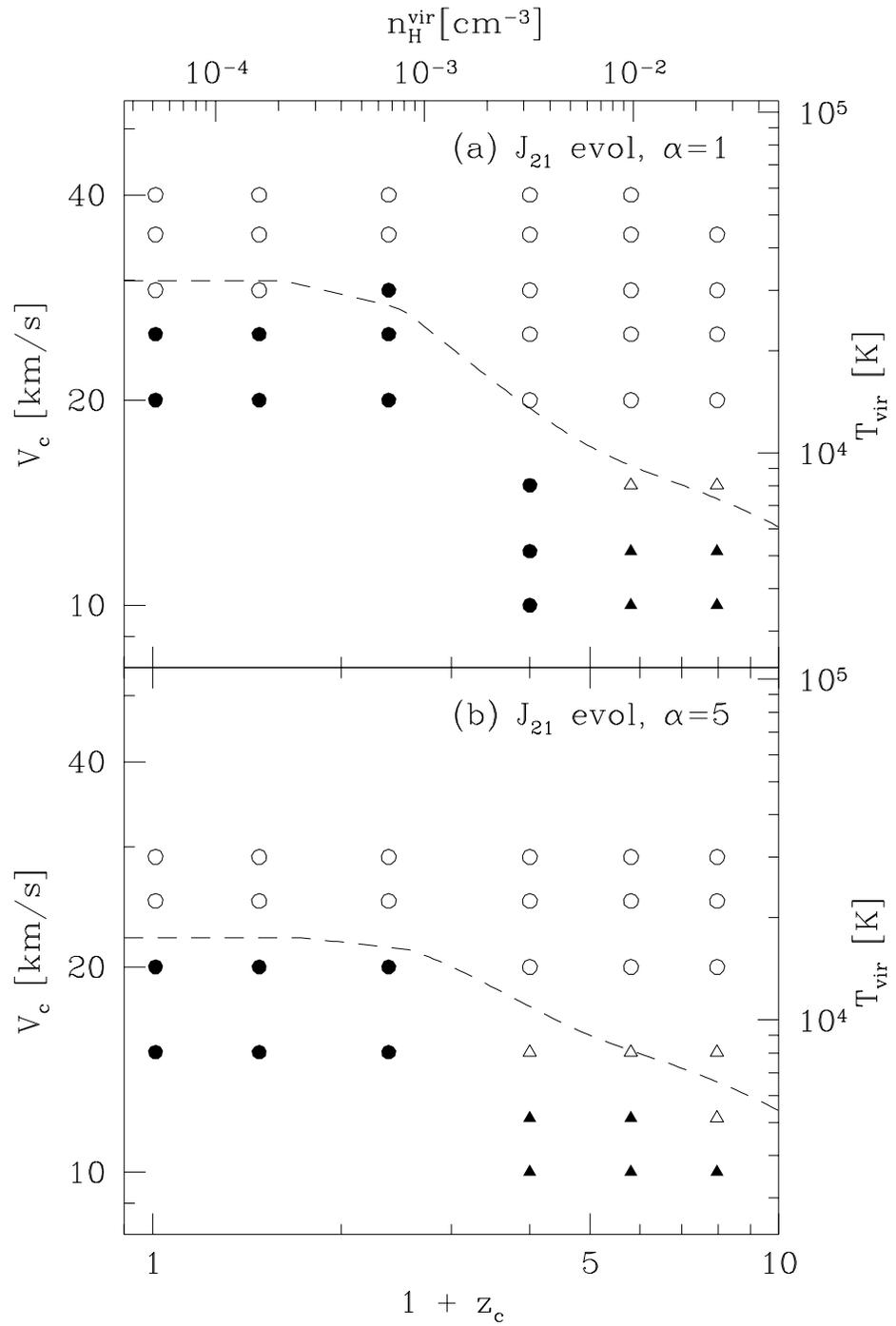,width=20cm}
\end{center}
\caption{Same as Fig.~\protect\ref{fig:vzc}\protect~ except that the
UV evolution is taken into account.}
\label{fig:vze}
\end{figure}

\end{document}